\documentclass[12pt]{iopart}

\usepackage{cite}
\usepackage{braket}
\usepackage{comment}
\usepackage{amssymb}
\usepackage{graphicx}
\usepackage{booktabs}
\newcommand\headercell[1]{%
	{\begin{tabular}[t]{@{}c@{}} #1 \end{tabular}}}
	
\begin{document}
\newcommand{\meh}{m$E$\textsubscript{h}}
\newcommand{\eh}{$E$\textsubscript{h}}

\title{Implementation of the Projective Quantum Eigensolver on a Quantum Computer}

\author{Jonathon P. Misiewicz and Francesco A. Evangelista}

\address{Department of Chemistry and Cherry Emerson Center for Scientific Computation, Emory University, Atlanta, GA 30322, U.S.A.}
\ead{francesco.evangelista@emory.edu}
\vspace{10pt}
\begin{indented}
\item[]October 2023
\end{indented}

\begin{abstract}
We study the performance of our previously proposed Projective Quantum Eigensolver (PQE) on IBM's quantum hardware in conjunction with error mitigation techniques. For a single qubit model of H\textsubscript{2}, we find that we are able to obtain energies within 4 millihartree (2.5 kcal/mol) of the exact energy along the entire potential energy curve, with the accuracy limited by both stochastic error and inconsistent performance of the IBM devices. We find that an optimization algorithm using direct inversion of the iterative subspace can converge swiftly, even to excited states, but stochastic noise can cause large parameter updates.
For the four-site transverse-field Ising model at the critical point, PQE with an appropriate application of qubit tapering can recover 99\% of the correlation energy, even discarding several parameters. The large number of CNOT gates needed for the additional parameters introduces a concomitant error that, on the IBM devices, results in loss of accuracy, despite the increased expressivity of the trial state. Error extrapolation techniques and tapering or postselection are recommended to mitigate errors in PQE hardware experiments.
\end{abstract}

\vspace{2pc}

\maketitle

\section{Introduction}

The rapid evolution of quantum hardware has opened a venue for addressing challenging problems in quantum chemistry with quantum computation \cite{Cao.2019.10.1021/acs.chemrev.8b00803,  Bauer.2020.10.1021/acs.chemrev.9b00829, Cheng.2020.10.3389/fchem.2020.587143, McArdle.2020.10.1103/RevModPhys.92.015003, Head-Marsden.2021.10.1021/acs.chemrev.0c00620, Ollitrault.2021.10.1021/acs.accounts.1c00514, Bharti.2022.10.1103/RevModPhys.94.015004, Miessen.2022.10.1038/s43588-022-00374-2, Motta.2022.10.1002/wcms.1580, Ayral.2023.10.48550/arXiv.2305.01097}.
However, quantum computing is currently not a practical tool for electronic structure theory. The heart of the problem is that existing implementations of digital quantum computers are noisy, intermediate-scale quantum (NISQ) devices\cite{Preskill.2018.10.22331/q-2018-08-06-79, Motta.2022.10.1002/wcms.1580, Lau.2022.10.1007/s43673-022-00058-z} with a limited number of qubits, short decoherence times, and subject to several sources of error.
Even highly accurate state-of-the-art quantum algorithms require sophisticated error mitigation techniques to tame the noise inherent to NISQ devices\cite{Cai.2023.10.48550/arXiv.2210.00921, Endo.2021.10.7566/JPSJ.90.032001, Qin.2022.10.1088/1674-1056/ac7b1e, Huang.2023.10.1007/s11433-022-2057-y}.
The ultimate test for quantum algorithms is, therefore, based on their utility in producing highly accurate results on hardware.
Within the NISQ context, one of the most promising classes of quantum algorithms for near-term applications is based on the variational quantum eigensolver (VQE)\cite{Peruzzo.2014.10.1038/ncomms5213, Tilly.2022.10.1016/j.physrep.2022.08.003, Fedorov.2022.10.1186/s41313-021-00032-6}, a hybrid quantum-classical approach that variationally optimizes a trial state parameterized by a quantum circuit.
In recent years, there has been a growing interest in finding better alternatives to VQE for NISQ devices.\cite{Stair.2021.10.1103/PRXQuantum.2.030301, Huggins.2022.10.1038/s41586-021-04351-z, Wang.2015.10.1021/acsnano.5b01651, Li.2023.10.48550/arXiv.2308.01559, Chee.2022.10.48550/arXiv.2207.03949, Boyn.2021.10.1063/5.0074842, Smart.2022.10.1103/PhysRevA.105.022405, Smart.2021.10.1103/PhysRevLett.126.070504, Smart.2023.10.48550/arXiv.2303.00758, Kumar.2023.10.48550/arXiv.2212.01801, Imoto.2023.0.48550/arXiv.2305.15974, Teplukhin.2020.10.1038/s41598-020-77315-4, Teplukhin.2021.10.1038/s41598-021-98331-y, Teplukhin.2019.10.1021/acs.jctc.9b00402, Genin.2019.10.48550/arXiv.1901.04715, Tazhigulov.2022.10.1103/PRXQuantum.3.040318, Sun.2021.10.1103/prxquantum.2.010317, Motta.2019.10.1038/s41567-019-0704-4, Sokolov.2023.10.1103/PhysRevResearch.5.023174, Lin.2021.10.1103/PRXQuantum.2.010342, Yeter-Aydeniz.2020.10.1038/s41534-020-00290-1, Barison.2022.10.1103/PhysRevA.106.022404, Oftelie.2022.10.1103/PhysRevLett.129.130603} However, most new quantum algorithms are usually assessed via emulators, and hardware benchmarks are scarce.

In this work, we implement and examine the performance of the projective quantum eigensolver (PQE)\cite{Stair.2021.10.1103/PRXQuantum.2.030301} on quantum hardware.
PQE is a NISQ-focused hybrid quantum-classical algorithm that seeks to optimize a parameterized trial state by satisfying projections of the Schr\"{o}dinger equation onto a many-body basis.
PQE has been studied exclusively in the context of method development. These include selected variants of PQE\cite{Stair.2021.10.1103/PRXQuantum.2.030301}, which is part of the family of adaptive methods\cite{Grimsley.2019.10.1038/s41467-019-10988-2, Stair.2021.10.1103/PRXQuantum.2.030301, Ryabinkin.2020.10.1021/acs.jctc.9b01084, Yordanov.2020.10.1038/s42005-021-00730-0, Tang.2021.PRXQuantum.2.020310}.
Two follow-up studies have focused on reducing the number of CNOT gates in unitary coupled cluster trial states combined with PQE \cite{Magoulas.2023.10.1021/acs.jctc.2c01016,Magoulas.2023.10.48550/arXiv.2304.12870}. More recently, post-iterative corrections to PQE have been explored.\cite{Magoulas.2023.10.1021/acs.jpca.3c02781}.
Additionally, a formula used in PQE for asymmetric expectation values was employed by Filip and co-workers in the context of projective quantum Monte Carlo\cite{Filip.2023.10.48550/arXiv.2306.14540} and by Angelakis et al. for an MP2 variant\cite{Chee.2022.10.48550/arXiv.2207.03949}. 
Maitra and coworkers proposed a ``two-phase'' PQE\cite{Halder.2023.10.1063/5.0155009}. While the first phase is standard PQE, the second phase computes the values of the ``small parameters'' by machine learning from the first phase and thereby requires fewer quantum measurements. Of greater relevance to this article, a recent publication from Maitra and co-workers (which appeared while this manuscript was being finalized) applied pre-existing error mitigation techniques to PQE \textit{on noisy emulators}\cite{Halder.2023.10.1063/5.0166433}.

Although classical numerical simulations of quantum algorithms are useful, actual hardware may respond differently from simulations, even for device-specific models of the computer\cite{Russo.2023.10.1109/TQE.2023.3305232, Martin.2022.10.1103/PhysRevResearch.4.023190}, thus warranting its own discussion.
In the literature, the closest reference points are VQE computations, as hardware results on PQE have not yet been reported. Some papers have reported VQE ground state energies or properties\cite{Kumar.2019.10.48550/arXiv.1903.03454, Arute.2020.10.1126/science.abb9811, Huang.2022.10.1021/acs.jpclett.2c02381, Huang.2023.10.1021/acs.jctc.2c01119, Zhang.2022.10.48550/arXiv.2206.07907, Huang.2022.10.1103/PRXQuantum.3.010339, Kandala.2019.10.1038/s41586-019-1040-7, Yu.2023.10.1103/PhysRevResearch.5.013183, Gao.2021.10.1021/acs.jpca.0c09530, Lolur.2023.10.1021/acs.jctc.2c00807, Suchsland.2022.10.1103/PhysRevResearch.4.013165, Sennane.2023.10.1103/PhysRevA.107.012416, Rice.2021.10.1063/5.0044068, Kandala.2017.10.1038/nature23879, Gao.2021.10.1038/s41524-021-00540-6, Colless.2018.10.1103/PhysRevX.8.011021, Hempel.2018.10.1103/PhysRevX.8.031022, O'Malley.2016.10.1103/PhysRevX.6.031007, Egger.2023.PhysRevResearch.5.033159, Crippa.2021.10.3390/magnetochemistry7080117, Obrien.2022.10.48550/arXiv.2210.10799, Tilly.2020.10.1103/PhysRevA.102.062425, Yamamoto.2022.10.1103/PhysRevResearch.4.033110, Zhang.2022.10.1038/s41534-022-00599-z, McCaskey.2019.10.1038/s41534-019-0209-0, Nam.2020.10.1038/s41534-020-0259-3, Martin.2022.10.1103/PhysRevResearch.4.023190, Sagastizabal.2019.10.1103/PhysRevA.100.010302, Czarnik.2022.10.48550/arXiv.2204.07109, Maupin.2023.10.48550/arXiv.2307.07027, Barison.2022.10.1103/PhysRevA.106.022404, Zhao.2023.10.1038/s41534-023-00730-8, Urbanek.2020.10.1103/PhysRevA.102.022427, Goings.2023.10.48550/arXiv.2308.00667, Ganzhorn.2019.10.1103/PhysRevApplied.11.044092, Peruzzo.2014.10.1038/ncomms5213, Huo.2022.10.1103/PhysRevA.105.022427, Stenger.2023.10.1103/PhysRevA.107.062606, Shen.2017.10.1103/PhysRevA.95.020501, Feniou.2023.10.48550/arXiv.2306.17159, Mazzola.2019.10/1103/PhysRevLett.123.130501, Bentellis.2023.10.48550/arXiv.2305.07092}.
Other studies have used VQE as an ingredient of methods to compute excited state energies and properties\cite{Khan.2023.10.1063/5.0144680, Huang.2023.10.1021/acs.jctc.2c01119, Huang.2022.10.1103/PRXQuantum.3.010339, Ollitrault.2020.10.1103/physrevresearch.2.043140, Gao.2021.10.1038/s41524-021-00540-6, Colless.2018.10.1103/PhysRevX.8.011021, Tilly.2020.10.1103/PhysRevA.102.062425, Yamamoto.2022.10.1103/PhysRevResearch.4.033110, Ma.2020.10.1038/s41524-020-00353-z, Barison.2022.10.1103/PhysRevA.106.022404, Ganzhorn.2019.10.1103/PhysRevApplied.11.044092}, linear response properties\cite{Huang.2022.10.1021/acs.jpclett.2c02381, Gocho.2023.10.1038/s41524-023-00965-1}, molecular dynamics\cite{Sokolov.2021.10.1103/PhysRevResearch.3.013125}, and vibrational eigenstates\cite{Lotstedt.2021.10.1103/PhysRevA.103.062609, Lotstedt.2022.10.1116/5.0091144}, while yet other studies have used VQE as an active space solver for a dynamical correlation or embedding method\cite{Krompiec.2022.10.48550/arXiv.2210.05702, Kirsopp.2022.10.1002/qua.26975, Tammaro.2023.10.1021/acs.jpca.2c07653, Rossmannek.2023.10.1021/acs.jpclett.3c00330, Tilly.2021.10.1103/PhysRevResearch.3.033230, Ma.2020.10.1038/s41524-020-00353-z, Rungger.2020.10.48550/arXiv.1910.04735, Backes.2023.10.1103/PhysRevB.107.165155, Kawashima.2021.10.1038/s42005-021-00751-9}.
Of particular interest to this paper are those VQE computations that report the effectiveness of different error mitigation schemes\cite{Zhang.2022.10.48550/arXiv.2206.07907, Khan.2023.10.1063/5.0144680, Huang.2022.10.1103/PRXQuantum.3.010339, Tammaro.2023.10.1021/acs.jpca.2c07653, Gao.2021.10.1021/acs.jpca.0c09530, Arute.2020.10.1126/science.abb9811, Lolur.2023.10.1021/acs.jctc.2c00807, Tilly.2021.10.1103/PhysRevResearch.3.033230, Gao.2021.10.1038/s41524-021-00540-6, Egger.2023.PhysRevResearch.5.033159, Obrien.2022.10.48550/arXiv.2210.10799, McCaskey.2019.10.1038/s41534-019-0209-0, Czarnik.2022.10.48550/arXiv.2204.07109, Maupin.2023.10.48550/arXiv.2307.07027, Urbanek.2020.10.1103/PhysRevA.102.022427, Huo.2022.10.1103/PhysRevA.105.022427, Bentellis.2023.10.48550/arXiv.2305.07092}, or that are performed on IBM devices\cite{Huang.2023.10.1021/acs.jctc.2c01119, Sokolov.2021.10.1103/PhysRevResearch.3.013125, Krompiec.2022.10.48550/arXiv.2210.05702, Lotstedt.2021.10.1103/PhysRevA.103.062609, Kumar.2019.10.48550/arXiv.1903.03454, Zhang.2022.10.48550/arXiv.2206.07907, Huang.2022.10.1103/PRXQuantum.3.010339, Kirsopp.2022.10.1002/qua.26975, Tammaro.2023.10.1021/acs.jpca.2c07653, Lotstedt.2022.10.1116/5.0091144, Ollitrault.2020.10.1103/physrevresearch.2.043140, Yu.2023.10.1103/PhysRevResearch.5.013183, Rossmannek.2023.10.1021/acs.jpclett.3c00330, Suchsland.2022.10.1103/PhysRevResearch.4.013165, Tilly.2021.10.1103/PhysRevResearch.3.033230, Rice.2021.10.1063/5.0044068, Gao.2021.10.1038/s41524-021-00540-6, Egger.2023.PhysRevResearch.5.033159, Crippa.2021.10.3390/magnetochemistry7080117, Tilly.2020.10.1103/PhysRevA.102.062425, Zhang.2022.10.1038/s41534-022-00599-z, McCaskey.2019.10.1038/s41534-019-0209-0, Martin.2022.10.1103/PhysRevResearch.4.023190, Gocho.2023.10.1038/s41524-023-00965-1, Czarnik.2022.10.48550/arXiv.2204.07109, Backes.2023.10.1103/PhysRevB.107.165155, Barison.2022.10.1103/PhysRevA.106.022404, Urbanek.2020.10.1103/PhysRevA.102.022427, Ganzhorn.2019.10.1103/PhysRevApplied.11.044092, Huo.2022.10.1103/PhysRevA.105.022427, Stenger.2023.10.1103/PhysRevA.107.062606, Mazzola.2019.10/1103/PhysRevLett.123.130501, Bentellis.2023.10.48550/arXiv.2305.07092}.
Since our computations will utilize IBM devices and performance is device-specific,\cite{Kordzanganeh.2023.10.1002/qute.202300043, Tilly.2020.10.1103/PhysRevA.102.062425, Bentellis.2023.10.48550/arXiv.2305.07092} comparisons with results from similar hardware are more apt.

Non-VQE methods have been applied to systems of chemical interest on hardware as well. Quantum devices have been used for ground state methods including quantum Monte Carlo\cite{Huggins.2022.10.1038/s41586-021-04351-z}, quantum phase estimation\cite{Wang.2015.10.1021/acsnano.5b01651}, perturbation theory\cite{Li.2023.10.48550/arXiv.2308.01559, Chee.2022.10.48550/arXiv.2207.03949}, and the Hermitian and anti-Hermitian contracted Schr{\"o}dinger equations\cite{Boyn.2021.10.1063/5.0074842, Smart.2022.10.1103/PhysRevA.105.022405, Smart.2021.10.1103/PhysRevLett.126.070504, Smart.2023.10.48550/arXiv.2303.00758}. Quantum annealing techniques have been used for ground states and other problems\cite{Kumar.2023.10.48550/arXiv.2212.01801, Imoto.2023.0.48550/arXiv.2305.15974, Teplukhin.2020.10.1038/s41598-020-77315-4, Teplukhin.2021.10.1038/s41598-021-98331-y, Teplukhin.2019.10.1021/acs.jctc.9b00402, Genin.2019.10.48550/arXiv.1901.04715}. Reduced density matrices from non-VQE methods may be passed into other, larger algorithms\cite{Boyn.2021.10.1063/5.0074842, Huang.2023.10.1103/PRXQuantum.4.020313}. Hardware experiments with variants of quantum imaginary time evolution (QITE)  have been used to query zero-temperature and finite-temperature properties on hardware\cite{Tazhigulov.2022.10.1103/PRXQuantum.3.040318, Sun.2021.10.1103/prxquantum.2.010317, Motta.2019.10.1038/s41567-019-0704-4, Sokolov.2023.10.1103/PhysRevResearch.5.023174, Lin.2021.10.1103/PRXQuantum.2.010342, Yeter-Aydeniz.2020.10.1038/s41534-020-00290-1, Barison.2022.10.1103/PhysRevA.106.022404, Oftelie.2022.10.1103/PhysRevLett.129.130603}. Hardware has also been used to study the dynamics of both closed\cite{Maruyoshi.2023.10.1088/1751-8121/acc369, Chen.2023.10.48550/arXiv.2302.10436, Sopena.2021.10.1088/2058-9565/ac0e7a, Lee.2022.10.1021/acs.jctc.1c01296, Lee.2023.10.1007/s40042-023-00722-z, Arute.2020.10.48550/arXiv.2010.07965, Zhukov.2018.10.1007/s11128-018-2002-y, Babukhin.2020.10.1103/PhysRevA.101.052337, Rost.2021.10.48550/arXiv.2108.01183, Smith.2019.10.1038/s41534-019-0217-0, Barends.2015.10.1038/ncomms8654, Gibbs.2022.10.1038/s41534-022-00625-0, Martin.2023.10.48550/arXiv.2305.19231, Doronin.2021.10.1007/s00723-021-01435-x, Oftelie.2020.10.1103/PhysRevB.101.184305, Tornow.2022.10.1088/1751-8121/ac6bd0, Lotstedt.2023.10.1088/1402-4896/acbcac, Lamm.2018.10.1103/PhysRevLett.121.170501, Urbanek.2021.10.1103/PhysRevLett.127.270502, Kim.2023.10.1038/s41586-023-06096-3} and open\cite{Guimaraes.2023.10.48550/arXiv.2302.14592, Guimaraes.2022.10.48550/arXiv.2203.14653, Peetz.2023.10.48550/arXiv.2307.14325, Head-Marsden.2021.10.1103/PhysRevResearch.3.013182, Gaikwad.2022.10.1103/PhysRevA.106.022424, Hu.2020.10.1038/s41598-020-60321-x, Han.2021.10.1103/PhysRevLett.127.020504, Kamakari.2022.10.1103/PRXQuantum.3.010320, Schlimgen.2021.10.1103/PhysRevLett.127.270503} quantum systems. Other algorithms without VQE have also been applied to hardware to study vibrational and molecular dynamics\cite{Richerme.2023.10.1021/acs.jpclett.3c01601, Gaidai.2022.10.1038/s41598-022-21163-x}, response properties\cite{Sun.2023.10.48550/arXiv.2302.04271}, eigenstates and electronic spectra\cite{Neill.2021.10.1038/s41586-021-03576-2, Zhu.2022.10.1088/1367-2630/ac61d1, Motta.2023.10.1039/d2sc06019a, Zhu.2021.10.1103/PhysRevA.103.032606, Santagati.2018.10.1126/sciadv.aap9646, Libbi.2022.10.1103/PhysRevResearch.4.043038, Cruz.2023.10.1103/PhysRevA.108.012618, Li.2019.10.1103/PhysRevLett.122.090504}, and even problems on the frontiers of interest to computational chemistry\cite{O'Brien.2022.10.1103/PRXQuantum.3.030345, Yoshida.2022.10.1063/5.0086489, Smart.2019.10.1103/PhysRevA.100.022517}.

The present article ties these two threads of PQE and hardware together, reporting the first realization of PQE on NISQ devices, with the goal of assessing the effectiveness of techniques to mitigate noise in the measurements and whether the algorithmic differences between PQE and VQE matter on hardware. The paper is organized as follows: Section \ref{sec:bg} provides a self-contained introduction to the theory of the PQE and a brief overview of the error mitigation techniques that we employ. The results of our study are found in Section \ref{sec:results}. We begin with a careful analysis of H\textsubscript{2} in a minimal basis set, described in Section \ref{subsec:h2}. After summarizing the encoding of the problem onto a quantum computer, we evaluate the noise in PQE at a fixed geometry due to finite measurement, compute an entire dissociation curve including a model of the quantum noise, and then compute a dissociation curve computed using IBM quantum devices. We then proceed to study the transverse-field Ising model (TFIM) in Section \ref{subsec:tfim}. We begin by validating our procedure, including a custom tapering procedure, on a noiseless simulator. We proceed to study the four-site problem, first with noise models and then with results from PQE simulations on quantum hardware. Afterwards, we compute TFIM with up to seven sites using noise models on quantum emulators, to compare PQE and VQE. We conclude with remarks on the prospects of PQE in Section \ref{sec:closing}.

\section{Background}
\label{sec:bg}

\subsection{Projective Quantum Eigensolver}
\label{sec:pqe}

Here, we summarize the PQE algorithm as introduced previously\cite{Stair.2021.10.1103/PRXQuantum.2.030301}. The ground-state energy of a quantum mechanical system is the lowest eigenvalue of its Hamiltonian, $H$. Let $\ket{\Phi_0}$ be a reference state in the domain of $H$, usually chosen as simultaneously an unentangled qubit register and close to an exact eigenstate of $H$, and let $U$ be a unitary operator on the domain of $H$. Then $U \ket{\Phi_0}$ is an eigenvector of $H$ if and only if $\ket{\Phi_0}$ is an eigenvector of $U^{-1} H U = \bar{H}$. This occurs when the projection of $\bar{H} \ket{\Phi_0}$ onto any vector $\ket{\Phi^\perp_0}$ orthogonal to $\ket{\Phi_0}$ is zero, i.e.,  $\braket{\Phi^\perp_0 | \bar{H} |  \Phi_0} = 0$. If we construct a basis for the orthogonal complement of $\ket{\Phi_0}$ it suffices to satisfy a \textit{residual equation} for each basis element.
\begin{equation}
\label{eq:general-r}
r_\mu \equiv \bra{\Phi_\mu} \bar{H} \ket{\Phi_0} = 0.
\end{equation}
When the residual equations are satisfied, the energy eigenvalue $E$ can then be read as
\begin{equation}
E = \bra{\Phi_0} \bar{H} \ket{\Phi_0}.
\end{equation}

In practice, we can neither enforce all residual equations nor vary across all possible unitaries $U$. Instead, we parameterize our unitary $U$ as a product of $N$ operators

\begin{equation}
    U(\vec{\theta}) = \prod_{\mu = 1}^{N} \exp(\theta_\mu \kappa_\mu)
\end{equation}
for real parameters $\theta_\mu$ and anti-Hermitian operators $\kappa_\mu$. We then define the orthogonal complement basis as the set of states

\begin{equation}
	\label{eq:phimu}
	\ket{\Phi_\mu} = \kappa_\mu \ket{\Phi_0}.
\end{equation}
As long as the $\ket{\Phi_\mu}$ are orthogonal to each other and to $\ket{\Phi_0}$, we then enforce the residual equation for each $\ket{\Phi_\mu}$. For fermionic systems, these $\kappa_\mu$ are generators of rotations between occupied (hole) and unoccupied (particle) orbital clusters in second quantization\cite{McArdle.2019.10.1103/physrevlett.122.180501}. If we denote occupied (unoccupied) orbitals with the indices $i,j,\ldots$ ($a,b,\ldots$), then the $\kappa_\mu$ operators are of the form $a_a^\dagger a_i - a_i^\dagger a_a, a_a^\dagger a_b^\dagger a_j a_i - a_i^\dagger a_j^\dagger a_b a_a$, etc., and are {in one-to-one correspondence with excited Slater determinants. Permuting two orbitals within a Slater determinant produces the same determinant, but with a phase of $-1$\cite{McArdle.2019.10.1103/physrevlett.122.180501}. We resolve the phase ambiguity by putting orbitals in a lexicographical order, which means that equation \ref{eq:phimu} requires a negative sign for some $\mu$. This carries through all equations involving $\ket{\Phi_\mu}$. We neglect the sign for clarity of exposition. For spin systems, these $\kappa_\mu$ are imaginary multiples of Pauli strings and are associated with a specific spin state on each site.

To ensure that the residuals are zero, we first need to measure them. In our previous paper\cite{Stair.2021.10.1103/PRXQuantum.2.030301}, we proposed a simple way to measure residual elements. Observe that for our choice of $\kappa_\mu$ we have that $\kappa_\mu \ket{\Phi_\mu} = - \ket{\Phi_0}$. Then $\exp(\theta_\mu \kappa_\mu) \ket{\Phi_0} = \cos(\theta_\mu) \ket{\Phi_0} + \sin(\theta_\mu) \ket{\Phi_\mu}$. We then showed that the real component of the residual element may be evaluated as 
\begin{equation}
    \label{eq:real-residual-3}
    r_\mu = E_\mu(\pi / 4) - \frac{1}{2} (E_\mu(\pi / 2) + E_\mu(0))
\end{equation}
using the expectation value
\begin{equation}
    E_\mu(\phi) = \bra{\Phi_0} \exp(-\phi \kappa_\mu) \bar{H} \exp(\phi \kappa_\mu) \ket{\Phi_0} .
\end{equation}
It is thus possible to measure a single residual element with two quantum measurements \textit{beyond} what is needed for the ground state energy $E = E_\mu(0)$, by analogy to the parameter-shift rule\cite{Schuld.2019.PhysRevA.10.1103/PhysRevA.99.032331, Vidal.2018.10.48550/arXiv.1812.06323, Kottmann.2021.10.1039/D0SC06627C}. Unlike the parameter shift rule, this formula requires three energies. We recently derived a residual formula more similar to the parameter-shift rule, which expresses the real component of the residual using only two energies:
\begin{equation}
    \label{eq:real-residual-2}
    r_\mu = \frac{1}{2} [E_\mu(\pi / 4) - E_\mu(- \pi / 4)] .
\end{equation}
We refer to Eq.~\ref{eq:real-residual-2} as the \textit{reference-shift rule}.
While these formulae are formally equivalent, it does not follow that they are equally performant on NISQ devices. Our initial numerical tests were not able to determine a statistically significant difference between Eqs.~\ref{eq:real-residual-3} and \ref{eq:real-residual-2} that was robust to changes in the number of shots and the noise model. Therefore, for aesthetic reasons, our computations will use equation \eref{eq:real-residual-2}.

In PQE, the parameters are then varied so that the measured residuals are zero, within some numerical tolerance. In general, the zeros will not be unique: not only may multiple $\vec{\theta}$ produce the same $U(\vec{\theta}) \ket{\Phi_0}$, but different $U(\vec{\theta}) \ket{\Phi_0}$ can approximate different eigenstates of $H$. Both of these hold for VQE as well.
For the optimization algorithm, we employ a quasi-Newton step accelerated with the direct inversion of the iterative subspace (DIIS) method\cite{Pulay.1980.10.1016/0009-2614(80)80396-4}, as proposed in our previous publication\cite{Stair.2021.10.1103/PRXQuantum.2.030301}. The use of machine learning to predict the values of small parameters from the large parameters has been studied elsewhere\cite{Halder.2023.10.1063/5.0155009}.

\subsection{Error Mitigation Techniques}
\label{ssec:err-mit}
We now sketch the error mitigation techniques we employ. For details, see Reference \citenum{Motta.2022.10.1002/wcms.1580} and references therein. The technical details of how these techniques were applied to individual systems are found in each system's respective subsection within our results.

\subsubsection{Readout Error Mitigation}

A major source of error in NISQ devices arises in the measurement process itself, i.e., readout error. For example, a qubit in the $\ket{1}$ state is mistakenly identified as being in the $\ket{0}$ state and vice versa. We correct for it by using $X$ gates to prepare the quantum register in a given bitstring and then measuring the occupation of all qubits. By repeating this operation for all bitstrings, we construct a linear transformation from \textit{actual} bitstring probabilities to \textit{experimentally observed} bitstring probabilities. When we obtain any experimental measurement probabilities, we use this matrix to estimate the bitstring probabilities in the absence of readout error via Qiskit's least squares procedure, which ensures that measurement probabilities stay between $0$ and $1$.

\subsubsection{Qubit Tapering}
\label{sssec:taper}

Qubit tapering is typically considered a technique to reduce the number of qubits, not the error. However, tapering may reduce the resources required, preventing the error from those resources. For example, in Section \ref{subsec:tfim}, a generator that entangles four qubits is replaced with one that entangles three qubits, and three parameters no longer require CNOT gates. It is hence legitimate to consider it as an error-mitigation technique.

We now discuss how to perform qubit tapering\cite{Bravyi.2017.10.48550/arXiv.1701.08213, Setia.2020.10.1021/acs.jctc.0c00113}. First, identify a Pauli string that commutes with the molecular Hamiltonian. When these exist, it is normally a consequence of some physically meaningful symmetry, e.g., total spin or a point group symmetry. It is possible to perform a similarity transformation so that the string becomes a single Pauli gate on qubit $n$. Upon performing this transformation, all operators in the Hamiltonian will take an $I$ or $Z$ in that qubit, and the eigenstate can be chosen to consist only of bitstrings with a $0$ for that qubit, or only of bitstrings with a $1$ for that qubit. The experimenter can then decide which choice is correct for the eigenstate of interest to them. The qubit is then eliminated from the computation, and the $I$ and $Z$ are replaced with their corresponding $+1$ or $-1$, depending on what that gate would return on qubit $n$. Although specific forms of this similarity transformation have been prescribed previously, those prescriptions are not essential to the method.

\subsubsection{Postselection}

When symmetry is present, some bitstrings should have no counts. Yet errors can introduce counts, and it is precisely this that postselection eliminates\cite{Bonet-Monroig.2018.10.1103/physreva.98.062339, McArdle.2019.10.1103/physrevlett.122.180501}. In particular, suppose a state is an eigenfunction of Pauli string $P$. If it is possible to simultaneously measure $P$ and other Pauli strings (such as those in the Hamiltonian), then when we measure the Pauli strings simultaneously, we can discard all measurements that yield incorrect expectation values for $P$.

We mention one technical limitation. Although quantum mechanics permits us to simultaneously measure any operators that mutually commute, current quantum hardware only permits us to simultaneously measure any operators that commute for every qubit in the Pauli string, i.e., that mutually qubitwise commute. To apply postselection, we must therefore perform a unitary transformation so all operators of interest qubitwise commute, as has been studied previously\cite{Gokhale.2019.10.48550/arxiv.1907.13623, Crawford.2021.10.22331/q-2021-01-20-385, Yen.2020.10.1021/acs.jctc.0c00008, Hamamura.2020.10.1038/s41534-020-0284-2} and must be resolved for any hardware application.

\subsubsection{Error Extrapolation}

Error extrapolation\cite{Li.2017.10.1103/PhysRevX.7.021050, Temme.2017.PhysRevLett.119.180509, Endo.2018.10.1103/PhysRevX.8.031027} is flexible in what error it corrects for, only asking that it can be scaled. The error strength is represented by $\epsilon$, and we seek the exact energy, which is obtained at $\epsilon = 0$. We do this by assuming the dependence of the energy on $\epsilon$ follows some user-provided functional form and fitting its parameters according to measurements at increased values of $\epsilon$. The ``error-free'' expectation value can then be read as the value at $\epsilon = 0$. Much like the complete basis set extrapolations of quantum chemistry\cite{Feller.2011.10.1063/1.3613639}, this method is inherently heuristic but can be quite reliable. However, error extrapolation ceases to be reliable for time-dependent noise\cite{Schultz.2022.PhysRevA.106.052406}.

A standard source of error targeted with error extrapolation is that coming from CNOT gates. This is justified when CNOT gates are the only qubit-entangling gates, and gates that act only on a single qubit have much smaller errors. In this case, we scale the noise by a factor of $2n +1$ by adding $n$ CNOT pairs, which are just the identity in an errorless circuit, to each CNOT gate.

To perform error extrapolation, it remains to choose the functional form of the extrapolation function. Although the original proposals of error extrapolation used a linear function\cite{Li.2017.10.1103/PhysRevX.7.021050, Temme.2017.PhysRevLett.119.180509}, exponential extrapolation was also found to achieve good accuracy \cite{Endo.2018.10.1103/PhysRevX.8.031027}. We shall study both, but advise the reader that we modified the exponential extrapolation procedure for numerical stability. Details are provided in \ref{sec:exp-appendix}. We also point out that Maitra and coworkers\cite{Halder.2023.10.1063/5.0166433} studied error extrapolation, but exclusively on simulators.

\section{Results}
\label{sec:results}

Unless otherwise indicated, all simulations on quantum hardware used 8,192 shots for each measurement.

\subsection{H\textsubscript{2}}
\label{subsec:h2}

\subsubsection{Simulation Details}
\label{sec:h2-sim-details}

We simulated H\textsubscript{2} with the minimal STO-6G atomic basis, which includes one basis function per atom. We first parameterize our state. From the atomic basis two spatial orbitals can be formed, a bonding orbital $\phi_g$ of $\sigma_g^+$ symmetry and an antibonding orbital $\phi_u$ of $\sigma_u^-$ symmetry. The ground state possesses $^1\Sigma_g^+$  symmetry, so by standard symmetry arguments, we can write it as:
\begin{equation}
\ket{\Psi} = c_g \ket{\phi_g \alpha \phi_g \beta} + c_u \ket{\phi_u \alpha \phi_u \beta} .
\end{equation}
We encode the two-dimensional space of all possible symmetry-allowed states onto a single qubit by mapping $\ket{\phi_g \alpha \phi_g \beta} $ to $\ket{0}$ and $\ket{\phi_u \alpha \phi_u \beta}$ to $\ket{1}$, so
\begin{equation}
    \ket{\Psi} = \cos(\theta/2) \ket{0} + \sin(\theta/2) \ket{1} = R_y(\theta) \ket{0} .
\end{equation}
The same ansatz may be derived by beginning with fermionic second quantized operators\cite{Helgaker.2000.10.1002/9781119019572} and applying Z2 symmetry reduction to the Jordan--Wigner encoding\cite{Bravyi.2017.10.48550/arXiv.1701.08213, Setia.2020.10.1021/acs.jctc.0c00113}. We immediately see that $U(\theta) = R_y(\theta)$. This may be efficiently implemented with no multi-qubit gates. With the encoding complete, we can then construct the Hamiltonian in the basis of our two states, using integrals obtained from the \textsc{Psi4} package\cite{Smith.2020.10.1063/5.0006002}.

We compute a dissociation curve from 0.4 to 6.0 \AA, and for each bond length, we perform ten hardware experiments. A hardware experiment consists of both the optimization of the amplitude to within a residual of $0.03$ and the measurement of the energy at the optimized amplitude. (We find that tighter convergence metrics significantly increases the number of iterations required with no appreciable increase in accuracy.) As the bond stretched, in some experiments we observed qualitatively incorrect energies. When these occurred, the energy points were excluded and re-computed. We discuss the origin of this effect below. We used a single readout error calibration matrix for each set of ten energy computations. All computations were performed within an hour of computing the readout error calibration matrix. Our initial guess was always $\theta = 0$.

\subsubsection{Results}
\label{sec:h2-results}

We show our H\textsubscript{2} dissociation curve in Figure \ref{fig:h2_diss_curve}, computed on IBM's Lima device. We observe that for all points, the averaged measured energy (shown as a red horizontal line) agrees with the observed energy to within four \meh, a good agreement considering the convergence threshold used.

Both the standard deviation of the data points and the error in the curve are smallest for points closest to dissociation and largest when the atoms are closest together, especially beyond the equilibrium point. Our group has observed a similar phenomenon previously\cite{Huang.2023.10.1103/PRXQuantum.4.020313}. We attribute this to the simple fact that the error in a term in the energy is directly proportional to a prefactor derived from matrix elements. The energy may be expressed in terms of Hamiltonian matrix elements $(H_{ij})$ and the nuclear repulsion energy $(E\textsubscript{nuc})$ as
\begin{equation}
 E = \frac{H_{00} + H_{11}}{2} + E\textsubscript{nuc} + H_{01} \braket{X} + \frac{H_{00} - H_{11}}{2} \braket{Z}
\end{equation}
While the prefactor for the $X$ string varies from 0.17 at compressed geometries to 0.34 at stretched geometries, the prefactor for the $\langle Z \rangle$ string varies from --0.99 to 0.00 along the same curve, being largest when the atoms are near each other. We therefore attribute both the larger variance and larger error at compressed geometries to the geometry dependence of the prefactor of the $Z$ term. (The magnitude of this effect, however, is larger than would we expected solely by comparing coefficients of $\langle Z \rangle$.) This stochastic error is the dominant effect for stretched geometries.

However, statistical error alone cannot account for the persistent fact that for compressed geometries, the energy frequently falls \textit{below} full configuration interaction. As we will discuss below, for this system, the PQE residual is equivalent to the VQE gradient, so we can regard this as a VQE optimization. We observe that noise effectively distorts the angle used for measuring $X$ relative to the angle used to measure $Z$, and the VQE uses this to drive the energy below full configuration interaction.

We caution the reader that due to the larger error in the three most compressed geometry points, we needed to re-compute these data points several times. We observed variations in energy of over four \meh\, even when no parameters were changed. In the final data we present, we magnified the number of shots used for readout calibration by a factor of 10, which should improve the accuracy of measuring the readout calibration matrix by a factor of $\sqrt{10}$\cite{McClean.2016.10.1088/1367-2630/18/2/023023}. After making this change, our energies improved by several \meh.

\begin{figure}
\begin{center}
\includegraphics{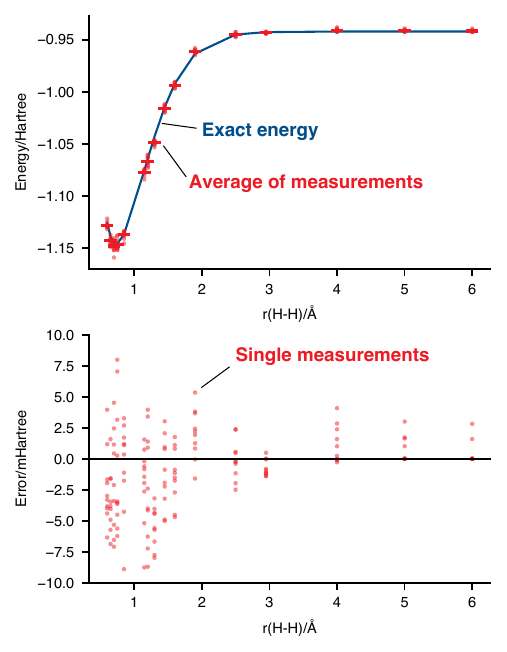}
\caption{Hardware experiments on the singlet ground state of the H\textsubscript{2} molecule in a minimal basis set. 
Top: Energies measured in experiments on a quantum computer (red) compared to exact full configuration interaction (blue curve) energies. Dots represent individual measurements, and horizontal lines indicate averages.
Bottom: Errors from full configuration interaction. All hardware experiments were performed on the IBM Lima device with 8,192 shots, readout error mitigation, and a 1-qubit encoding.}
\label{fig:h2_diss_curve}
\end{center}
\end{figure}

Because PQE is an iterative algorithm, a complete cost estimate must consider the number of iterations required to converge. We show this for two representative geometries in Figure \ref{fig:h2_conv_curve}. When the bond distance is 0.75 \AA, close to equilibrium and where the initial guess is quite accurate, energy convergence is rapid, while amplitude convergence takes until after three iterations. At 2.95 \AA, where the initial guess has only 54\% overlap with the exact ground state, convergence is much more erratic but relatively swift, being achieved within eight iterations. The reader may also note a large spread in the energies after two rounds of parameter optimization, with some energies worse than the initial guess. This is an artifact of our optimization algorithm. A pure Jacobi step will only cause a large amplitude update when the residual or approximate Hessian is large. Neither is the case here. When we optimize with DIIS, the optimization is sensitive to stochastic noise in the measurements and can cause very large extrapolations. Despite the energy having a period of 2$\pi$ in the amplitude, we have observed cases where the DIIS optimizer sends parameter values far away from the initial guess of 0, into the neighborhood of $\pm 1000$. We have observed similar behavior on a noiseless simulator, but still found the performance of DIIS to be superior to the performance of a pure Jacobi step. We have found that taking the DIIS amplitude modulo the period eliminates this behavior and leads to swifter convergence on noisy simulators, but have not tested this on quantum hardware. Other modifications to the DIIS update, such as a trust radius, can also be imagined.

\begin{figure}
\begin{center}
\includegraphics{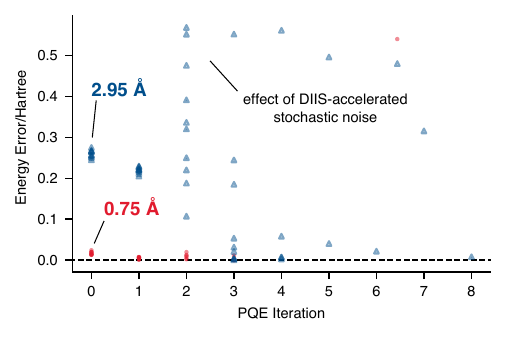}
\caption{Hardware experiments on the singlet ground state of the H\textsubscript{2} molecule in a minimal basis set.
Convergence of the energy error with respect to full configuration interaction after $n$ iterations of the PQE algorithm for ten runs. All hardware experiments were performed on the IBM Lima device with 8,192 shots, readout error mitigation, and a 1-qubit encoding.}
\label{fig:h2_conv_curve}
\end{center}
\end{figure}

We noted in Section \ref{sec:h2-sim-details} that for stretched geometries, the PQE algorithm returned qualitatively incorrect energies, in error by over 0.5 \eh. We were able to reproduce this behavior on a quantum simulator with no noise model and shall henceforth discuss results on the simulator, where we could run PQE repeatedly. We found that the erroneous energies were within 5 millihartree of the energy of the \textit{excited-state} at that geometry. To understand this, recall that we initialize the wavefunction in the state $\ket{0}$. As the bond stretches, the overlap of $\ket{0}$ with the ground state diminishes from over 99\% to 50\%. As our optimizer tries to zero projections, which can be done by targeting any eigenstate, it is unsurprising in this case stochastic errors direct the optimization toward the excited state.

We view this result positively: in PQE, with an appropriate choice of amplitudes, it is possible to target an excited state. (We refer the interested reader to other quantum algorithms\cite{Mizuta.2021.10.1103/physrevresearch.3.043121, Jones.2019.10.1103/physreva.99.062304, Parrish.2019.10.1103/physrevlett.122.230401, Ollitrault.2020.10.1103/physrevresearch.2.043140, Higgott.2019.10.22331/q-2019-07-01-156, Yeter-Aydeniz.2020.10.1038/s41534-020-00290-1, Tsuchimochi.2023.110.1021/acs.jctc.2c00906, Libbi.2022.10.1103/PhysRevResearch.4.043038} to compute excited states.) Indeed, by changing our initial guess, we have been able to target the excited state dissociation curve on the quantum simulator. The utility of this feature of PQE depends on how easy it is to obtain guess amplitudes and how accurately higher solutions of an approximate unitary ansatz can represent an excited state. This is a promising avenue of future work. 

Lastly, we compared the performance of PQE to that of VQE after each optimization iteration, at the minimum energy geometry. The results are shown in Figure \ref{fig:h2_pqe_vs_vqe}. The numerical performance is nearly identical. In fact, it may be shown that for the special case of one-parameter systems, the PQE residual is the VQE gradient. (This is most easily shown by comparing the two-measurement formulas for the residual to the VQE shift rule.) At least in this case, all difference between the two algorithms is due to numerical noise and does not indicate a meaningful performance difference in the algorithms.

\begin{figure}
\begin{center}
\includegraphics{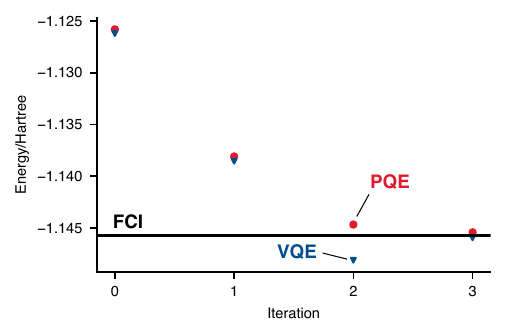}
\caption{
Simulation of hardware experiments on the singlet ground state of the H\textsubscript{2} molecule in a minimal basis set.
Comparison of the energy convergence averaged over ten runs of the PQE and VQE algorithms.
All simulations utilized the error model for the IBM Lima device, employing 8,192 shots and a 1-qubit encoding.}
\label{fig:h2_pqe_vs_vqe}
\end{center}
\end{figure}

\subsection{Transverse-Field Ising Model}
\label{subsec:tfim}

\subsubsection{Simulation Details}

The Hamiltonian for the transverse-field Ising model (TFIM) on $N$ sites is given by

\begin{equation}
    \label{eq:tfim-h}
    H = - h \sum\limits_{i = 1}^N Z_i + J \sum\limits_{i = 1}^{N-1} X_i X_{i+1} \quad ,
\end{equation}
which describes a 1-D chain of spin sites subject to two kinds of interaction. First, adjacent spin sites have a co-alignment-penalizing interaction with magnitude controlled by $J$. Second, an external electric or magnetic field interacts with all sites, with magnitude controlled by $h$, rewarding sites that align with the field. All Pauli strings in $H$ commute with the Pauli string $\prod_{i=1}^N Z_i$, which may be physically interpreted as reversing the direction of the field and mathematically interpreted as requiring that for any eigenstate, the bitstrings all have the same number of $\ket{1}$ qubits, modulo two, i.e., parity symmetry. In the context of the present work, this symmetry enables postselection and the tapering of one qubit.

The present work chooses $h = J = 1$, and, aside from the last subsection, we choose $N = 4$. With this choice of parameters, we have modeled a strongly correlated system. The choice of $J / h = 1$ puts this system at the critical point where it transitions between ferromagnetic and antiferromagnetic phases in the thermodynamic limit\cite{He.2017.10.1088/1742-5468/aa85b0, Lieb.1961.10.1016/0003-4916(61)90115-4}. In the uncorrelated basis of $\{\ket{\uparrow}, \ket{\downarrow}\}^{\otimes N}$, the overlap of any basis state with the correlated wavefunction is at most 81.2\%, and the correlation energy (defined as the difference between the TFIM solution with $J = 0$ and the exact energy)} is 15.9\% of the total energy.

We have chosen our wavefunction ansatz as the so-called ``quantum combinatorial ansatz'' of Reference \citenum{Herasymenko.2021.10.22331/q-2021-12-02-596}. The wave operator is a product of exponentiated imaginary multiples of Pauli strings. Each Pauli string is a product of a single $Y$ and $X$ gates, which always acts on the largest index of the non-trivial indices for that string. This ansatz is known to represent any state with real wavefunction coefficients (for complex coefficients, strings without a $Y$ gate must also be allowed). We loop over the largest qubit index, from smallest to largest, but operator order is otherwise arbitrary. In some particular cases, we have chosen to neglect operators which are unlikely to have large amplitudes. We shall explicitly say when we do this. An advantage of this choice of ansatz is that it is an exact ansatz for any state, tapered or not. By symmetry, we may set to zero the coefficients of all operators in the untapered ansatz with an \textit{odd} number of non-trivial operators. We choose our starting state to be $\ket{0}^{\otimes N}$, and this choice of operators preserves the parity symmetry.

Let us discuss how symmetry applies to the use of postselection. Here, there are two sets of mutually commuting operators to measure. The first set consists of $Z$ operators on various qubits, all of which already commute qubitwise with $\prod_{i=1}^N Z_i$. The second set consists of $X_i X_{i+1}$, which does not commute qubitwise with $\prod_{i=1}^N Z_i$. To transform these operators into ones that can be simultaneously measured, we prepend a CNOT staircase, a sequence of CNOT gates where qubit $i$ controls qubit $i+1$ for all qubits. This transforms $X_i X_{i+1}$ into $X_{i+1}$, allowing for the simultaneous measurment of all such operators\cite{Sun.2021.10.1103/prxquantum.2.010317}. Our use of tapering is more involved and is described in the next section.

\subsubsection{Exact Results}

Before analyzing the robustness of PQE to noise, we first study our algorithm on a noiseless simulator. With the procedure deployed above, we are able to reproduce the exact energy, as determined by Hamiltonian diagonalization, to within machine precision. This confirms the basic correctness of our ansatz and of PQE, when noise and noise mitigation are neglected. Looking at the corresponding parameters, we observe that they fall into two classes. One consists of all coefficients of $X_i Y_{i+1}$ Pauli strings, which all have magnitude of at least $0.45$. The second class of parameters have a smaller value, with magnitude less than $0.12$. Such a partitioning of PQE parameters was central to Reference \citenum{Halder.2023.10.1063/5.0155009} and could also have been anticipated from the perturbative arguments of Reference \citenum{Herasymenko.2021.10.22331/q-2021-12-02-596}.
If these three large parameters are optimized while all other parameters are set to zero, we are able to retain 96\% of the correlation energy. The remainder of this section is concerned with how this division can be exploited in conjunction with qubit tapering.

As discussed in Section \ref{sssec:taper}, standard qubit tapering requires a choice of qubit to be tapered out. When we attempt all four qubit taperings from the standard procedure, we are able to reproduce the exact energy, but find that only tapering qubits one or two provides an ansatz that retains at least 96\% of the correlation energy when only three parameters are optimized. The other choices of qubits to taper lead to only 86\% of the correlation energy when only the three largest parameters are retained. The exact solutions for the other two taperings blur the separation into large and small parameters. Tapering qubit three yields one parameter that increases to magnitude $0.35$, and tapering qubit four yields two parameters that increase to magnitude $0.23$ and $0.27$, respectively. But even the two more accurate taperings are less than satisfactory. For one, the resulting Hamiltonians include a term that acts on \textit{every qubit}, which costs the locality of the Hamiltonian. Additionally, the parameters that are important still retain at least one generator that acts on multiple qubits and is, therefore, more likely to introduce additional noise in the experiments. It would be best if the large parameters were associated with action on only \textit{one} qubit.

We identified a non-standard tapering procedure that achieves this goal. Any parity-allowed basis state can be reached by applying a sequence of $X_{i} X_{i+1}$ operations to the starting state, $\ket{0}^{\otimes 4}$. Starting from the first qubit, apply $X_i X_{i+1}$ if needed to achieve the target value on qubit $i$, and use parity to enforce that the final qubit is correct. We map each basis state to a state where index $i$ is $1$ if $X_i X_{i+1}$ was needed, and $0$ otherwise, producing an overall state of $3$ qubits. For example, $\ket{0101} = (X_0 X_1) (X_1 X_2) \ket{000}$ maps to $\ket{011} = X_0 X_1 \ket{000}$. We can then construct an operator in this reduced space that produces the same expectation values in the old space, after accounting for the transformation. Notice that the operator $X_i X_{i+1}$ will always transform to $X_i$ in this alternative tapering scheme.

We have confirmed that this tapering scheme reproduces the exact energy. As expected, there are three large parameters of value at least $0.45$, corresponding to the $X_i$, while all other parameters are of magnitude below $0.12$. Restricting the ansatz to the ``large'' parameters leads to recovering 97\% of the correlation energy, as desired. Further, all terms in the Hamiltonian involve either one qubit or two adjacent qubits, meaning locality is restored.

\begin{figure}
\begin{center}
	\includegraphics[width=\textwidth]{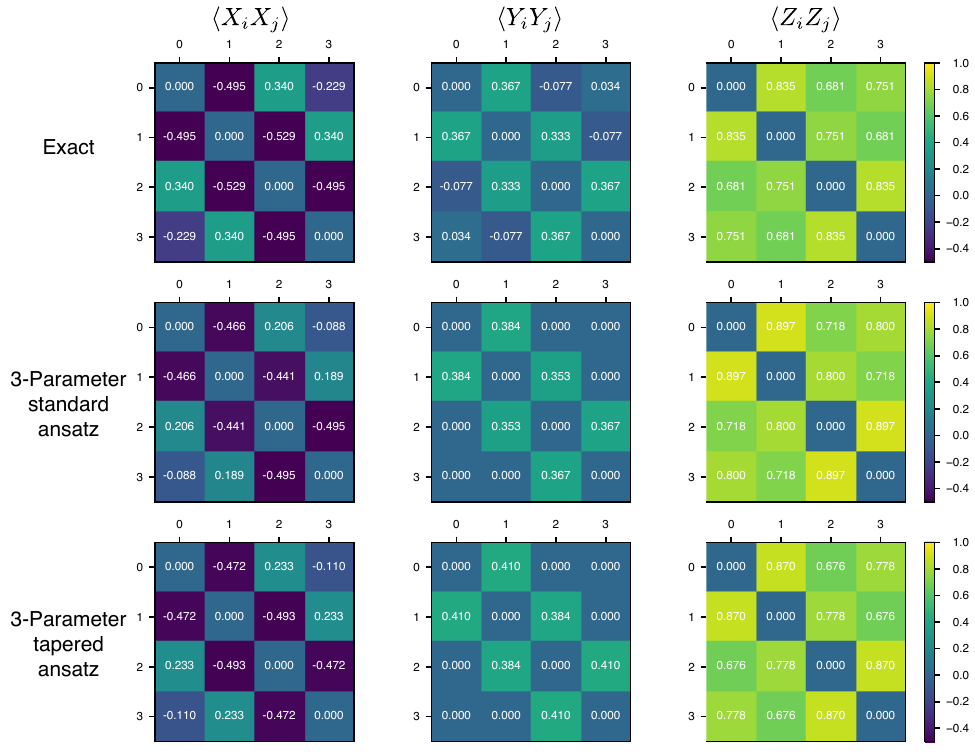}
	\caption{Spin-spin correlation functions of the 4-site transverse-field Ising model. The left, center, and right columns measure correlation in the X, Y, and Z directions, respectively. The top row uses full configuration interaction, the middle row uses the three largest parameters in the standard ansatz, and the bottom row uses the three largest parameters in our tapered ansatz.}
	\label{fig:tfim}
\end{center}
\end{figure}

Before proceeding, let us assess the quality of the wavefunctions produced by eliminating small parameters. We can do this by measuring the spin-spin correlation functions, $\braket{\sigma_i \sigma_j}$, which measure the alignment of spin in the direction $\sigma \in \{X, Y, Z\}$ on qubits $i$ and $j$. In Figure \ref{fig:tfim}, we see that many of the correlation functions are quite close, but those correlation functions between non-adjacent qubits, whose generators have been set to zero, can be far off. The same trend continues for the YY and ZZ correlation functions, albeit less strongly. We see similar behavior for our custom tapering scheme, which tends more accurate for XX and ZZ correlation, but slightly less accurate for YY correlation. We caution that we need to transform the $\sigma_i \sigma_j$ operators into the new basis, which we do not discuss here. Nonetheless, we see results of similar quality. This confirms that our representation of the wavefunction has not been impaired by tapering.

\subsubsection{Simulator Results}
\label{ssec:tfim-simulator-4}

We begin our exploration of the transverse-field Ising model by studying combinations of the error mitigation techniques described in Section \ref{ssec:err-mit} using a simulator of the noisy Lima device of IBM. One instance of our simulations are displayed in Table \ref{tbl:tfim-sim}. We caution the reader that due to both the stochastic nature of quantum sampling and, more importantly, drift in the error model describing the device, repeating this experiment leads to different percentages, with greater difference from 100\% leading to greater variability. For example, the correlation energy recovered can vary between 67\% and 78\% when postselection is used without extrapolation. Drift in the noise model leads us to recover between --68\% and --28\% of the correlation energy when only readout error is mitigated.

	\begin{table}[h!]
	\begin{center}
	    \caption{The percentage correlation energy recovered for the transverse-field Ising model on a simulator modeling noise from IBMQ's Lima device. Readout error mitigation was used for all computations. Results are averaged over 100 runs.}
		\begin{tabular}{ cccc }
			\hline 
			
			\hline 			
			\headercell{Symmetry Use} & \multicolumn{3}{c@{}}{Extrapolation Method}\\
			\cmidrule(l){2-4}
			 & None & Linear & Exponential\\ 
			\hline
			None & $-68$\% & $36$\% & $97$\% \\
			Tapering & $19$\% & $82$\% & $98$\%\\
			Postselection & $67$\% & $98$\% & $104$\% \\
			\hline 
			
			\hline
		\end{tabular}
	\label{tbl:tfim-sim}
	\end{center}	
	\end{table}

What is qualitatively clear is that for the 4-site TFIM, the extrapolation techniques are ordered, as are the symmetry exploiting techniques. Postselection is more accurate than qubit tapering, which is more accurate than merely removing symmetry-breaking generators. Exponential extrapolation is more accurate than linear extrapolation, which is more accurate than no extrapolation. Although it is useful to learn that eliminating erroneous measurements via postselection is a better way to eliminate error than eliminating extra qubits via tapering, all other comparisons of methods may have been easily anticipated.

\subsubsection{Hardware Data}

We now turn to experiments on quantum hardware. We found that the noise on IBM devices fluctuated in intensity to the point that even with error mitigation, we would observe qualitatively different results from experiments submitted just hours apart. Significant hardware drift over time has also been reported previously\cite{Czarnik.2022.10.48550/arXiv.2204.07109}. In one instance, results that ran and completed within an hour of each other differed by $-0.2$ \eh\ due to IBM recalibrating their hardware between the experiments. This severely limited the rigor with which we could perform experiments. We also observed that when we attempted error extrapolation, for some circuits, the error upon adding additional CNOT gates led the count distribution to be \textit{further away} from pure noise, violating the assumptions of the exponential error extrapolation model discussed in \ref{sec:exp-appendix}. Accordingly, exponential error extrapolation could not be performed reliably. Degraded performance of error extrapolation on hardware compared to a noise model on an emulator has been reported previously\cite{Roggero.2020.10.1103/PhysRevC.102.064624}, and appropriate forms of error mitigation have been known to be hardware-dependent\cite{Maupin.2023.10.48550/arXiv.2307.07027}. For other uses of extrapolation schemes on hardware, see references \citenum{McArdle.2021.10.1103/PRXQuantum.2.020349, Giurica-Tion.2020.10.1109/QCE49297.2020.00045, Huang.2022.10.1103/PRXQuantum.3.010339}.

One particularly successful experiment used our custom tapering for the 4-site TFIM and included \textit{only} the three non-entangling generators, i.e., using an approximate ansatz that can be evaluated more accurately on the quantum hardware. We repeated this ten times and measured a final energy of $-4.77 \pm 0.03$ \eh. The exact energy is $-4.76$ \eh. Within the precision possible on NISQ devices, the accuracy cannot reasonably be improved. An extension of this tapering scheme to molecular systems is unclear. Regardless, these findings make clear that for qubit tapering, some similarity transformations vastly outperform others.

We report additional experiments that are encouraging for PQE: we combined our tapering scheme, with all generators except the three-qubit entangler, with exponential extrapolation. On the Lima device, this reproducibly gave 120\% of the correlation energy. By combining linear extrapolation and qubit tapering on the IBM Perth device, as well as removing the three-site generator, we were able to measure an energy of $-4.70$ \eh\, i.e., we recovered 92\% of the correlation energy. Given this data point and the similarity of the PQE and VQE equations, we suspect that the difficulties we observed measuring data from PQE are not specific to the PQE algorithm but reflect the challenges of obtaining reliable measurements from NISQ devices with all but the simplest quantum circuits. We add that although we attempted simulations with postselection, the results were inferior to those of exponential extrapolation and tapering.

\subsubsection{PQE and VQE Comparison}
While the hardware results of the preceding section indicate strong noise effects, they do not conclusively establish whether the poor performance is unique to PQE. To investigate that question, we return to the simulators and perform PQE and VQE computations on the transverse-field Ising model with increasing sites. We employ IBM's noise model of their Perth device, since the Lima device used earlier in the paper has only five qubits, but our experiments require seven. Numerical experiments on the Lima device, for those systems where it was possible, show the same PQE and VQE comparison as do the Perth systems, but with a shift in the amount of correlation energy recovered that increases with the number of qubits. We employ linear and exponential error mitigation, with both tapering and postselection, as Section \ref{ssec:tfim-simulator-4} showed these strategies to be advantageous. To minimize stochastic finite measurement error, we magnified our standard 8,192 shots by a factor of 40. We performed VQE using analytic gradients augmented by the shift rule, accelerated by DIIS. For additional discussion on combining VQE and DIIS, see \ref{sec:diis-appendix}.

\begin{figure}
     \includegraphics[width=\textwidth]{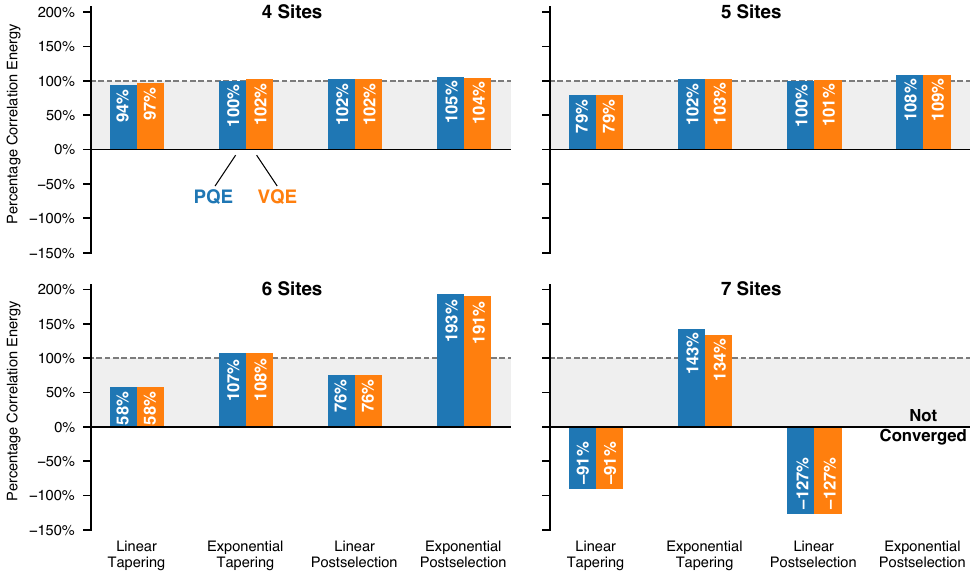}
     \caption{
Simulation of hardware experiments on the transverse-field Ising model with 4 to 7 sites.    
     Correlation energy measured employing all combinations of postselection or tapering, with linear or exponential error extrapolation.
All simulations utilized the error model for the IBM Perth device, employing 327,680~shots and a 1-qubit encoding.}     
     \label{fig:pqe-vs-vqe-tfim}
\end{figure}

The results are shown in Figure \ref{fig:pqe-vs-vqe-tfim}. PQE and VQE results are extremely close in all cases. They disagree the most, by percentage, when exponential extrapolation is combined with tapering for the 7-site problem. With this combination of error mitigation techniques, although individual computations converged, different simulations would lead to different energies, and we had to repeat the experiment many times to obtain converged statistics. We suspect that performing \textit{more} computations would yield closer agreement. Nonetheless, even in this case, it is clear that the main reason for the poor quality of our hardware results is due to hardware noise and not PQE or the optimization algorithm employed.. Rather, the noise is too intense for either PQE OR VQE using analytic residuals or gradients, respectively.

From Figure \ref{fig:pqe-vs-vqe-tfim}, we can also observe that error steadily increases as the number of qubits increases, from 4 to 7. For the 7-qubit simulations, combining exponential extrapolation and postselection does not yield converging energies. This matter is discussed in \ref{sec:exp-appendix}. Combining exponential extrapolation with tapering is the most robust approach for the larger systems, but even in this best-case scenario, the error is already prohibitive at 7 qubits. Although the conclusion from Section \ref{ssec:tfim-simulator-4} that exponential extrapolation is better than linear extrapolation still holds, the preference for postselection reverses once exponential extrapolation is consistently employed.

\section{Conclusions}
\label{sec:closing}

In this article, we have studied the performance of the Projective Quantum Eigensolver (PQE) on quantum hardware, to assess whether it is a viable and competitive algorithm to solve for molecular electronic structure on NISQ devices. While we encountered difficulties in applying PQE, these can be attributed to the inherent challenges of using NISQ devices, even with error mitigation. We applied PQE to H\textsubscript{2} in a minimal basis set with qubit tapering and mapped the state to a one qubit problem. In this case, the PQE residual condition is formally identical to zeroing gradients of VQE computed via the parameter-shift rule. For this problem, PQE converges swiftly near the equilibrium geometry. However, when the initial guess is not as accurate, PQE may either converge to an excited state or become more sensitive to errors, leading to a large spread of energies. For the four-site transverse-field Ising model (TFIM), PQE performs excellently on a noisy simulator, recovering over 99\% of the correlation energy using a combination of exponential error extrapolation and postselection. On physical hardware, errors complicate the picture. While we have managed to recover over 90\% of the correlation energy for the TFIM, device errors prevent us from doing this consistently. We found more success by using a custom qubit tapering that allows for 99\% accuracy to be achieved with generators that require no CNOT gates. More reliable hardware is required before PQE and VQE can be definitively compared.

We now identify directions for future development of the PQE, until devices become more robust to error. We have identified new formulae for the residuals that need to be measured in the theory. Although not discussed in great detail here, the similarities to the shift rule allow for more fruitful adaptation of ideas from selected VQE into selected PQE.
Second, we identified potential for improvement in the optimization of PQE. During the course of our studies, providing a suitable initial guess and exploiting periodicity in the parameters were both found to improve the convergence of DIIS. These areas merit further investigation, as do alternative convergence algorithms. Work in this direction has already begun\cite{Halder.2023.10.1063/5.0155009}. Lastly, and perhaps most intriguingly, our results on H\textsubscript{2} suggest that PQE can be used to target excited states, by identifying other roots of the PQE residual conditions and leveraging the fact that PQE does not use direct minimization. We plan to further explore these excited state PQE solutions in our subsequent research.

\section{Acknowledgments}
This work is supported by the U.S. Department of Energy under Award No. DE-SC0019374 and the NSF under Grant No. CHE-2038019. We acknowledge helpful discussions with Ilias Magoulas, Renke Huang, Harper Grimsley, Shi-ning Sun, and Austin Minnich. We acknowledge the use of IBM Quantum services for this work. The views expressed are those of the authors and do not reflect the official policy or position of IBM or the IBM Quantum team.

\appendix

\section{Exponential Error Extrapolation}
\label{sec:exp-appendix}

In exponential error extrapolation, the dependence of the expectation value of a Pauli string (other than the identity) is modeled as an exponential function of the error strength, $E(\epsilon) = \lambda \exp(\kappa \epsilon)$, where $\lambda$ and $\kappa$ are adjustable parameters. It follows that $E(0) = \lambda$, which can be computed from the values of $E(\epsilon)$ for two particular $\epsilon$. The formula specifically involves the ratio of the measured energies. Therefore, the exponential model inherently assumes:
\begin{itemize}
    \item The expectation value does not flip sign.
    \item The ratio of the expectation values is numerically stable.
    \item Random error, e.g., from sampling, is negligible in comparison to the systematic error being corrected.
    \item The expectation value decreases in magnitude as error increases, going to zero in the infinite noise strength limit
\end{itemize}

The first two of these assumptions will fail when the measured expectation values are near-zero, which will happen for many Pauli strings. The third assumption will fail when systematic noise only weakly affects a particular expectation value, which the fourth assumption predicts when the true expectation value is near-zero. Hence, in cases where we have small expectation values, exponential error extrapolation is unreliable due to sampling noise, but the fourth assumption predicts that it should have a small impact anyways. PQE likely has more of these small expectation values than VQE without analytic gradients, because PQE must measure residuals involving multiple wavefunctions that are not localized in Hilbert space.

Accordingly, for each Pauli string, we measured the expectation values for both noise strengths \textit{as if} we would perform exponential extrapolation and then checked that the expectation values had the same sign, decreased as error increased, that the high-error expectation was no smaller than 0.0001, and that the ratio of the errors was no greater than 2500. If any of those conditions were false, we did not perform exponential extrapolation and assumed that the low-noise value was sufficiently accurate: failure of these assumptions is likely when the true expectation value is already small. Otherwise, we performed exponential error extrapolation. Naively performing exponential error extrapolation in all cases led to numerical instabilities.

The two numerical thresholds mentioned in the preceding paragraph were chosen arbitrarily based on numerical experience with the 4-site transverse-field Ising model. We found these thresholds not to be robust for different systems. For the 7-site transverse-field Ising model, we frequently observed some cases where the error ratio was on the order of 1000, due to a large measurement on the first extrapolation point. This led to the already large measurement being amplified by a factor of $\sqrt{1000}$. This is fundamentally the reason we observe the non-convergence reported in Figure \ref{fig:pqe-vs-vqe-tfim}. While tighter thresholds would prevent error magnification, they fail to correct the error in the first data point.

\section{DIIS Acceleration of VQE}
\label{sec:diis-appendix}

Reference \citenum{Stair.2021.10.1103/PRXQuantum.2.030301} reported that VQE required 2 to 3 times as many gradient evaluations as PQE to obtain a given energy accuracy, at least for the system under consideration. We were not able to reproduce this in our experiments with PQE and VQE on transverse-field Ising models. While we found that PQE did converge faster than VQE in the absence of noise, this happened by one to two iterations.

To understand this difference, note that Reference \citenum{Stair.2021.10.1103/PRXQuantum.2.030301} compared PQE \textit{using DIIS} to VQE \textit{using the BFGS algorithm}\cite{Broyden.1970.10.1093/imamat/6.3.222, Fletcher.1970.10.1093/comjnl/13.3.317, Goldfarb.1970.10.1090/s0025-5718-1970-0258249-6, Shanno.1970.10.1090/s0025-5718-1970-0274029-x}. Our experiments used DIIS for both systems. Upon repeating the experiments of Reference \citenum{Stair.2021.10.1103/PRXQuantum.2.030301} with DIIS for both algorithms, we found that PQE required an iteration or two fewer than VQE, as in our results. We thus conclude that the reports of swifter convergence of PQE than VQE from Reference \citenum{Stair.2021.10.1103/PRXQuantum.2.030301} were incomplete. The convergence speed depends on the optimization algorithm. The two algorithms perform comparably with DIIS, with PQE having a slight edge. At least for minimal amounts of noise, VQE optimization with DIIS requires fewer gradients than with BFGS. We make no statement regarding other optimization algorithms or performance under more realistic noise models. Shortcomings of DIIS optimization were noted in Section \ref{sec:h2-results}, and the problem of PQE optimization was studied in Reference \citenum{Halder.2023.10.1063/5.0155009}.

\section*{References}
\bibliographystyle{unsrt}
\bibliography{bibliography}
\end{document}